\documentclass[aps,prd,groupedaddress,preprintnumbers]{revtex4}
\usepackage{dcolumn}
\usepackage{latexsym}
\usepackage{graphicx}
\usepackage{pstricks}
\usepackage{epsfig}
\usepackage{color}

\input epsf
\usepackage{amsmath,amssymb,dsfont,amsfonts}

\newcommand{\be}{\begin{equation}}
\newcommand{\ee}{\end{equation}}
\newcommand{\bea}{\begin{eqnarray}}
\newcommand{\eea}{\end{eqnarray}}

\newskip\humongous \humongous=0pt plus 1000pt minus 1000pt

\newif\ifdtup

\newcommand{\lapprox}{%
\mathrel{%
\setbox0=\hbox{$<$}
\raise0.6ex\copy0\kern-\wd0
\lower0.65ex\hbox{$\sim$}
}}
\newcommand{\gapprox}{%
\mathrel{%
\setbox0=\hbox{$>$}
\raise0.6ex\copy0\kern-\wd0
\lower0.65ex\hbox{$\sim$}
}}

\begin{document}


\title{The Muon Anomalous Magnetic Moment }

\author{Marc Knecht}
\email{knecht@cpt.univ-mrs.fr}
\affiliation{Centre de Physique Th\'{e}orique, CNRS/Aix-Marseille Univ./Univ. du Sud Toulon-Var (UMR 7332)\\
CNRS-Luminy Case 907, 13288 Marseille Cedex 9, France
}

\begin{abstract}
The calculations entering the prediction of the standard model value for the anomalous
magnetic moment of the muon $a_\mu$ are reviewed, and compared
to the very accurate experimental measurement.
The situation for the electron is discussed in parallel.\hfill

\end{abstract}

\maketitle

\section{Introduction}
\setcounter{equation}{0}

\noindent
The Dirac equation, together with minimal coupling to an external electromagnetic field,
predicts, for an elementary charged lepton $\ell$ of spin $1/2$, mass $m_\ell$, and charge $e_\ell$, 
a Pauli-type coupling ${\mbox{\boldmath $\mu_{\ell}$}}\cdot{\mbox{\bf B}}$ of the spin to the
magnetic field, where the magnetic
moment, proportional to the particle's spin, is given in units of the Bohr magneton, by
\begin{equation}
{\mbox{\boldmath $\mu_{\ell}$}}
=
g_{\ell}^{\rm Dirac}\,\left(\frac{e_{\ell}}{2m_{\ell}c}\right)\,
 \hbar\,\frac{\mbox{\boldmath$\sigma$}}{2}
 ,
\end{equation}
with $g_\ell^{\rm Dirac} = 2$.
Quantum corrections arising from loops will bring in new contributions and shift  
the gyromagnetic factor $g_\ell$ away from this value. Given
a sufficient level of precision, all degrees of freedom, known or unknown (i.e. light or
heavy), will eventually contribute in a visible way. The existence of an {\it anomalous} magnetic moment 
$a_\ell\equiv (g_\ell - g_\ell^{\rm Dirac})/g_\ell^{\rm Dirac}$ is thus an indirect probe 
of the existence and properties of all degrees of freedom, within, but also beyond, the 
standard model.

\noindent
It actually turns out that, on the experimental level, the anomalous magnetic moments of the 
electron ($a_e$) and of the muon ($a_\mu$) are among the most precisely measured low-energy 
observables in particle physics. In the case of the muon, the very precisely
known experimental value \cite{PDG12}
\begin{equation}
a_\mu^{\mbox{\footnotesize{exp}}} =  11 \, 659 \,208.9 (6.3) \cdot 10^{-10}\quad [0.54{\rm ppm}]
,
\end{equation}
is dominated by the results obtained by the BNL-E821 experiment \cite{Bennett06,Roberts09}.
The situation is even more impressive in the case of the electron. The value
\begin{equation}
\!\!\!\!\!
a_e^{\mbox{\footnotesize{exp}}} = 1 \,159 \, 652 \, 180 . 73 (0.28) \cdot 10^{-12}\ [0.24{\rm ppb}]
\end{equation}
follows from the measurement, at a relative 
precision of $0.28$ppt, of the gyromagnetic factor $g_e$ \cite{Hanneke08}.

\noindent
The high level of precision achieved by these experiments is certainly challenging for theory 
and theoreticians. The question naturally arises, whether theoretical evaluations are able to 
reach the same level of accuracy, and whether the standard model accounts for these 
values. Most exciting, of course, is the possibility that eventually a significant discrepancy (with
reliable theoretical and experimental uncertainties) remains, thus signaling the existence
of degrees of freedom beyond the standard model. 

\noindent
This review summarizes the present status of the theoretical evaluations of $a_\mu$ and of
$a_e$. In order to reach an accuracy comparable to the experimental results, it is necessary
to consider contributions from electromagnetic interactions (section 2), from weak interactions
(section 3), and from strong interactions (section 4). The latter, although larger than
the corrections from the standard model weak interactions, are particularly challenging,
since they heavily involve the non-perturbative regime. This is why they have been kept for the end. 
This review will end with Section 5, which includes a summary and some conclusions, and presents
perspectives for the future.

\section{QED contributions}

\noindent
The interactions involving only photons and charged leptons can be treated
perturbatively, 
\noindent
{\scriptsize
\begin{table}[t]
\setlength{\tabcolsep}{0.92pc}
 \caption{\scriptsize    The coefficients $C_\ell^{(2n)}$,
 of the perturbative QED expansion for
the anomalous magnetic moment of the electron (left) and
of the muon (right). The values are taken from \cite{QED_sum_ae}
and \cite{QED_sum_amu}.}
    {\small
\begin{tabular}{|c|c|c|} 
\hline 
   & $\ell = e$   &   $\ell = \mu$ \\
   \hline
$C_\ell^{(2)}$   &   $0.5$   &   $0.5$   \\
$C_\ell^{(4)}$   &   $-0.328 \, 478 \, 444 \, 00\ldots$   &   $0.765 \, 857 \, 425(17)$   \\
$C_\ell^{(6)}$   &   $1.181 \, 234 \, 017\ldots$   &   $24.050 \, 509 \, 96(32)$   \\
$C_\ell^{(8)}$   &   $-1.9096(20)$   &   $130.879 \, 6(63)$   \\
$C_\ell^{(10)}$   &   $9.16(58)$   &   $753.29(1.04)$   \\  
\hline
\end{tabular}
}
\label{table_1}
\end{table}
}
\noindent
\begin{equation}
a_{\ell}^{\rm QED} = 
\sum_{n\ge1} C_\ell^{(2n)} \left(\frac{\alpha}{\pi}\right)^n
.
\end{equation}
\noindent
The challenge comes, however, from the high  
orders in the perturbative expansion that one needs to consider in order
to reach a level of precision comparable to the experimental one. In the
case of the electron, the experimenal uncertainty 
$\Delta a_e^{\mbox{\footnotesize{exp}}} = 2.8 \cdot 10^{-13}$,
together with $(\alpha/\pi)^4 \sim 3 \cdot 10^{-11}$, indicates that one needs to 
compute at least five loop contributions. The one-loop coefficient
$C_e^{(2)}=C_\mu^{(2)} = 1/2$ was obtained by J. Schwinger 
\cite{Schwinger48} long ago. The two-loop 
and three-loop coefficients are also known analytically. 
For surveys of these calculations, and references to the original works 
see \cite{QED_calculations}.
The higher order coefficients involve 891 diagrams
at four loops, and 12672 diagrams at five loops. They have been computed
through systematic numerical evaluations of the multidimensional
integrals over the corresponding Feynman parameters.
Complete results have now been published \cite{4and5loops,QED_sum_ae,QED_sum_amu}. The results
of these calculations are displayed in Table \ref{table_1}, and we merely 
make a few comments.\\
{\bf i)}~The accuracy of the coefficients
$C_\mu^{(4)}$ and $C_\mu^{(6)}$, which are known analytically, is actually
limited by the uncertainties on the experimental values \cite{CODATA}
of the mass ratios $m_\mu/m_e$ and, to a lesser extent, $m_\mu/m_\tau$
and $m_e/m_\tau$. In the electron case, 
these uncertainties would only affect the digits beyond those shown, and are 
not relevant given the present size of $\Delta a_e^{\mbox{\footnotesize{exp}}}$.
The uncertainties on the four and five-loop coefficients come,
in both cases, from the numerical integration procedure.\\
{\bf ii)}~In the muon case, the experimental
accuracy is somewhat lower, $\Delta a_\mu^{\mbox{\footnotesize{exp}}} = 6.3 \cdot 10^{-10}$,
and one might think that even the fourth order is not needed. But this is
without reckoning with the structure of the coefficients, which, in the 
muon case, are logarithmically enhanced by the presence of the much lighter electron
in the loops. Starting at three loops, one encounters terms involving $\pi^2 \ln (m_\mu^2/m_e^2) \sim 50$!
This explains why the coefficients $C_\mu^{(2n)}$ in Table \ref{table_1} are typically 
larger than in the electron case. 
Despite the size of $C_\mu^{(10)}$, $C_\mu^{(10)}  (\alpha/\pi)^5 \sim 0.5 \cdot 10^{-10}$,
so that this contribution remains marginal in view of the present experimental 
error. But it will also have to be considered as the latter improves
(see Section 5).\\
{\bf iii)}~Some subsets of diagrams contributing to $C_\ell^{(8)}$ and $C_\ell^{(10)}$ are also
known analytically \cite{Broadhurst93,Laporta94,Aguillar08,Baikov08,Kurz14a}, 
in terms of expansions in powers of the mass ratios. These analytical results
provide useful and welcome checks of the numerical evaluations.\\
{\bf iv)}~Finally, notice that $\Delta C_\mu^{(2n)} (\alpha/\pi)^n$,
where $\Delta C_\mu^{(2n)}$ denotes the uncertainty on 
$C_\mu^{(2n)}$, is below the experimental
precision $\Delta a_\mu^{\mbox{\footnotesize{exp}}} = 6.3 \cdot 10^{-10}$ 
for $n=2,3,4,5$, so that in practice $a_{\mu}^{\rm QED}$ is, for the time being,
free of any theoretical error.  

At this stage, using the latest high-precision measurement
of the fine-structure constant \cite{Bouchendira11}
\begin{equation}
\alpha^{-1} = 137.035 \, 999 \, 037(91)
\quad [0.66{\rm ppb}]
,
\end{equation}
we have 
\begin{equation}
a_{e}^{\rm QED} =
1 \, 159 \, 652 \, 180.07(6) (4)(77) \cdot 10^{-12} ,
\end{equation}
and
\begin{equation}
a_{\mu}^{\rm QED} =
1 \, 165 \, 847 \, 189.51(19)(7)(77)(9) \cdot 10^{-12} .
\end{equation}
The first two errors come from the contributions at orders ${\cal O}(\alpha^4)$
and ${\cal O}(\alpha^5)$, respectively, and the third one comes from the
experimental uncertainty on $\alpha$. In the case of $a_{\mu}^{\rm QED}$
there is an additional error from the uncertainties on the mass ratios.
It then follows that
\begin{eqnarray}
a_e^{\mbox{\footnotesize{exp}}} - a_{e}^{\rm QED} &=& + 0.67 (82) \cdot 10^{-12}
,
\nonumber\\
a_\mu^{\mbox{\footnotesize{exp}}} - a_{\mu}^{\rm QED} &=& + 737.0 (6.3) \cdot 10^{-10}
.
\end{eqnarray}
In the case of the electron, $a_e$ is well described by QED, within the uncertainties, which are dominated
by the present uncertainty on the determination of $\alpha$.
In the case of the muon, the discrepancy with the experimental value is substantial,
so that the difference is to be ascribed to the contributions of the weak
and of the strong interactions.\\
Notice that one can use the measurement of $a_e$ in order to
reduce (by a factor 2.5) the error on $\alpha$ and obtain
a determination of the fine-structure constant at the $0.25$ppb
level \cite{QED_sum_ae}. But this means that $a_e$ then no longer provides
a possible test of the standard model (more on this at the end of Section 5).

\section{Contributions from the weak interactions}

\noindent
The one-loop contributions due to the weak interaction sector of the standard model
have been computed \cite{Jackiw72,Bars72,Fujikawa72,Altarelli72,Bardeen72} 
more than forty years ago. They read
\begin{equation}
a_{\ell}^{\rm weak(1)} =
\frac{G_F}{\sqrt{2}}\,\frac{m_{\mu}^2}{8\pi^2}\,
\Big[\frac{5}{3}+\frac{1}{3}
\left(1-4\sin^2\theta_{W}\right)^2 
+
{\cal O}\left(\frac{m_{\mu}^2}{M_{Z}^2}\log\frac{M_{Z}^2}{m_{\mu}^2}
\right) \! +
{\cal O}\left(\frac{m_{\mu}^2}{M_{H}^2}\log\frac{M_{H}^2}{m_{\mu}^2}
\right)
\bigg]
,
\end{equation}
and correspond numerically to $a_{\mu}^{\rm weak(1)} = 19.48\cdot 10^{-10}$.
Two-loop electroweak corrections are also available \cite{Czarnecki95,Knecht02b,Czarnecki03}, and reduce the
above value by about 20\%. A recent reevaluation \cite{Gnendinger13} gives, for the
sum of the one- and two-loop weak contributions,
\begin{eqnarray}
a_{\mu}^{\rm weak} &=& + 15.4(1) \cdot 10^{-10} 
,
\nonumber\\
a_{e}^{\rm weak} &=& + 0.297(5) \cdot 10^{-13} 
.
\end{eqnarray}

\noindent
After taking these contributions into account, we now have
\begin{equation}
a_\mu^{\mbox{\footnotesize{exp}}} - a_{\mu}^{\rm QED} - a_{\mu}^{\rm weak}
= + 721.65 (6.38) \cdot 10^{-10}
,
\end{equation}
and only the strong interactions remain 
in order to close this gap.

\section{Contributions from the strong interactions}

\noindent
The strong-interaction contributions mainly come
from the low-energy, light-quark, sector, so that a perturbative approach 
is no longer adapted. It is both useful and customary
to distinguish three types
of corrections involving the hadronic sector, as shown in Fig. \ref{fig-1}.
The diagram on the left corresponds to hadronic contributions to the
photon vacuum polarization function. The diagram in the middle
is the so-called hadronic light-by-light contribution. Finally,
there is also a hadronic contribution to the two-loop weak
corrections discussed in the previous section, arising from
the exchange of a virtual photon and a virtual neutral weak
gauge boson between the external lepton line and the hadronic
blob (right diagram). In comparison to the other two-loop weak
corrections, this last correction is small. It was already included in the
values given in the preceding section, and will not be discussed
further here. For details, I refer the reader to the literature
quoted in the preceding section.

\subsection{Hadronic vacuum polarization}

\noindent
Let us start by examining the case of hadronic vacuum polarization.
This contribution can be expressed as \cite{Bouchiat61,Durand62,Gourdin69}
\begin{equation}
a_{\ell}^{\rm HVP-LO}
=
\frac{1}{3}\,\left(\frac{\alpha}{\pi}\right)^2\,
\int_{4M_\pi^2}^{\infty} \,\frac{ds}{s}\,K(s) R^{\rm had} (s)
,
\label{HVP-LO}
\end{equation}
with
\begin{equation}
K(s) = \int_0^1 dx\,\frac{x^2(1-x)}{x^2 + (1-x)\,\frac{s}{m_{\ell}^2}}
,
\end{equation}
and $R^{\rm had} (s)$ represents the $R$-ratio of the cross section
of $e^+ e^- \to {\rm hadrons}$. The advantage of this representation
is threefold. First it tells us that the contribution is positive
($K(s)>0$).
Second, that it is dominated by the low-energy domain, since
$K(s)\propto m_\ell^2/s$ for $s$ large. And finally, it is directly related to an experimental
quantity, which allows $a_{\ell}^{\rm HVP-LO}$ to be evaluated using available
data on $e^+ e^- \to {\rm hadrons}$. The results from the two latest published 
evaluations \cite{Davier11,Hagiwara11}
give comparable values,
\begin{eqnarray}
a_{\mu}^{\rm HVP-LO} &=& +692.3(4.2) \cdot 10^{-10}  \quad [29]
,
\nonumber\\
a_{\mu}^{\rm HVP-LO} &=& +694.9(4.3) \cdot 10^{-10} \quad [30]
.
\end{eqnarray}
The corresponding value for the electron reads \cite{Nomura13}
\begin{equation}
a_{e}^{\rm HVP-LO} = +1.866(11) \cdot 10^{-12} 
,
\end{equation}
and is larger than the present uncertainty on 
$a_e^{\rm QED}$.

\begin{figure}[t]
\center\epsfig{figure=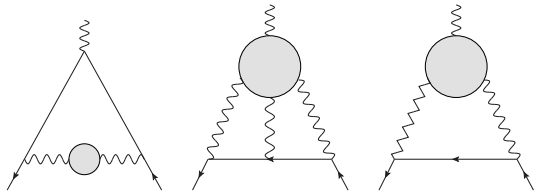,height=3.0cm} 
\caption{Hadronic contributions to the anomalous magnetic moment of
a charged lepton. One distinguishes hadronic vacuum polarization insertion
(left), involving the $\langle VV \rangle$ two-point function,
hadronic light-by-light scattering (middle), involving the $\langle VVVV \rangle$ 
four-point function, and a hadronic contribution to the two-loop weak corrections (right),
involving the $\langle VVA \rangle$ three-point function. 
Here $V$ stands for the hadronic part of the electromagnetic current, and $A$
denotes the axial part of the hadronic neutral current. In the last two cases,
only one typical diagram (out of 6 and 2, respectively) is displayed. At the
perturbative level, the blob would correspond to a quark loop, dressed with
additional virtual gluons and quarks.} \label{fig-1}
\end{figure}

\noindent
There are also contributions of the same type at order ${\cal O}(\alpha^3)$,
obtained upon inserting a second (hadronic or leptonic) 
vacuum polarization in
one of the two virtual photons of the leftmost diagram in Fig. \ref{fig-1},
or upon inserting a hadronic vacuum polarization in one of the
photon lines of the two-loop QED contributions. These contributions
can be written in a similar form as in Eq. (\ref{HVP-LO}), but involve
now a (known) kernel different from $K(s)$ \cite{Barbieri75,Krause97}. 
These next-to-leading
hadronic vacuum polarization corrections, evaluated with the same 
$e^+ e^- \to {\rm hadrons}$ data, amount to \cite{Hagiwara11}
\begin{equation}
a_{\mu}^{\rm HVP-NLO} = -9.84(7) \cdot 10^{-10}  
\end{equation}
for the muon, and \cite{Nomura13} $a_{e}^{\rm HVP-NLO} = -2.234(14) \cdot 10^{-13}$
for the electron. Recently, the next-next-to-leading
hadronic vacuum polarization corrections were also evaluated
\cite{Kurz14b},
with the result
\begin{equation}
a_{\mu}^{\rm HVP-NNLO} = +1.24(1) \cdot 10^{-10} 
.
\end{equation}
Summing these three contributions gives $a_{\mu}^{\rm HVP} = + 683.7(4.2) \cdot 10^{-10}$
with the input from \cite{Davier11},
or $a_{\mu}^{\rm HVP} = + 686.3(4.3) \cdot 10^{-10}$
with the input from \cite{Hagiwara11}, and the difference
$a_\mu^{\mbox{\footnotesize{exp}}} - a_{\mu}^{\rm QED} - a_{\mu}^{\rm weak} - a_{\mu}^{\rm HVP}$
equals  $+ 37.95(7.64) \cdot 10^{-10}$ or $+ 35.35(7.69) \cdot 10^{-10}$,
respectively.

\subsection{Hadronic light-by-light scattering}

\noindent
We have kept for the end the hadronic contribution
to virtual light-by-light scattering, that remains
the most challenging one at present. Here, we have 
no direct link to an experimental observable, and
even the overall sign cannot be fixed from the outset. 
One has to resort to a certain amount of model
dependence in order to describe, 
in the non-perturbative regime, the hadronic four-point
function (see Fig. \ref{fig-1}) that is involved. Only two
calculations have tried to give a complete
description of this four-point function from
the point of view of its contribution to $a_\mu$. After correcting
for the sign
\cite{Knecht02a} of the dominant contribution coming from the
reducible exchange of a single pseudoscalar meson between
pairs of (virtual) photons, their results
\cite{Bijnens95,Hayakawa95} read
\begin{eqnarray}
a_\mu^{\rm HLxL} &=& + 8.3(3.2) \cdot 10^{-10}
\quad [36],
\nonumber\\
a_\mu^{\rm HLxL} &=& + 8.96(1.54) \cdot 10^{-10}
\quad [37]
,
\end{eqnarray}
and show good agreement.
But if one looks into the details \cite{Prades}
of these calculations, there are
sometimes sizeable differences, either between
contributions common to the two models, or
in the contributions that are included or not.

\noindent
The limit of a large number of colours $N_c$
\cite{tHooft74,Witten79} can be useful in order to
organize the various contributions \cite{deRafael94}. 
At leading order, one has to consider contributions 
from reducible single meson
exchanges (pseudoscalars, axial vectors, tensors).
Of these, those of higher masses will have their
contributions suppressed, so that the contribution
from single $\pi^0$ exchanges dominates. This
is indeed the case in the models that have been
considered in \cite{Bijnens95,Hayakawa95}.
In the framework of the effective low-energy theory, the
leading-order contribution can be worked out
exactly \cite{Knecht02c},
\begin{equation}
 a_\mu^{{\rm HLxL}; \pi^0} = N_c \left( \frac{\alpha}{\pi} \right)^3
  \frac{N_c}{F_\pi^2} \frac{m_\mu^2}{48 \pi^2}
 \left[ \ln^2 \frac{M_\rho}{M_\pi}
 + c_\chi \ln \frac{M_\rho}{M_\pi} + \kappa
 \right]
 .
 \label{large-Nc}
\end{equation}
In this expression, $F_\pi$ is the pion-decay
constant ($F_\pi \sim \sqrt{N_c}$ in the large-$N_c$
limit), and $M_\rho$, the mass of the $\rho$ meson,
is a typical hadronic mass scale where the
effective theory ceases to be applicable.
The constant $c_\chi$ in front of the subleading
contribution is related \cite{Knecht02c,Ramsey02}
to a low-energy constant $\chi$ \cite{Savage92}
that also appears in the decay $\pi^0 \to e^+ e^-$,
and can thus be extracted \cite{Husek14} from data.
The last constant, $\kappa$, is not fixed by any
requirement.
In (\ref{large-Nc}), the sign of the leading contribution comes out as positive.
Since it results from a model-independent approach,
this result has a general validity.
In model calculations, the scale $M_\rho$ enters
through the form factors that describe the
$\pi^0 - \gamma^* - \gamma^*$ vertex, and which
are necessary in order to make the contribution
from $\pi^0$ exchange finite (taking the constant vertex that
follows from the Wess-Zumino term would lead to a divergent
contribution). As the UV-regulator $M_\rho$ is sent to infinity,
one should recover the behaviour (\ref{large-Nc}).
This has been checked in a variety of models \cite{Knecht02a}.

\noindent
Although the pion-exchange practically gives the final
result in \cite{Bijnens95,Hayakawa95}, this partly
also results from the cancellation among other contributions,
which, taken individually, can be sizeable. 
Some of these contributions are regularly updated
or reevaluated (for a recent example see \cite{Pauk14}, 
and the contributions in \cite{Mainz14}). There are
also contributions coming from short distances, and
an important issue is the matching between contributions
arising from different scales and regimes. This
is certainly a feature that one should improve.
For instance, although some short-distance constraints are
implemented \cite{Bijnens95,Melnikov04,Nyffeler09b}, an extensive study of
the four-point function (cf. Fig. \ref{fig-1})
from this point of view
has not been done so far. Eventually, the total error
on $a_\mu^{\rm HLxL}$ will not only reflect our ability
to reduce the errors
on individual contributions, but must also reflect
our confidence in the way we are able to put them
together, while respecting the known properties
of QCD at short and at long distances.

\noindent
In order to summarize the present situation,
one may quote the
``best estimate" from \cite{Prades10}
\begin{equation}
a_\mu^{\rm HLxL} = 10.5(2.6) \cdot 10^{-10}
,
\end{equation}
whereas a more conservative estimate \cite{Nyffeler09} gives
\begin{equation}
a_\mu^{\rm HLxL} = 11.5(4.0) \cdot 10^{-10}
.
\end{equation}
In the case of the electron, this contribution is much smaller,
and one finds \cite{Prades10}
\begin{equation}
a_e^{\rm HLxL} = 0.035(10) \cdot 10^{-12}
.
\end{equation}

\section{Summary, conclusions, perspectives}

\noindent
The anomalous magnetic moments of the electron and of the muon
are among the most precisely measured low-energy
observables of the standard model. 
The standard-model value obtained for $a_e$
agrees with the experimental measurement
\begin{equation}
 a_e^{\mbox{\footnotesize{exp}}} - a_e^{\rm SM} =
 -1.04(82) \cdot 10^{-12}
 .
\end{equation}
However, at present the value
obtained for $a_\mu$ wihin the standard model misses
the experimental one. The difference 
$a_\mu^{\mbox{\footnotesize{exp}}} - a_\mu^{\rm SM}$
ranges from $2.8\sigma$, taking the inputs
from \cite{Hagiwara11} for HVP and from \cite{Nyffeler09} for HLxL,
\begin{equation}
a_\mu^{\mbox{\footnotesize{exp}}} - a_\mu^{\rm SM} = 23.7(8.6)
\cdot 10^{-10}
\quad [30,~51]
,
\end{equation}
to $3.4\sigma$,
\begin{equation}
a_\mu^{\mbox{\footnotesize{exp}}} - a_\mu^{\rm SM} = 27.4(8.0)
\cdot 10^{-10}
\quad [29,~50]
,
\end{equation}
when the inputs for HVP and HLxL are taken from
\cite{Davier11} and \cite{Prades10}, respectively.
This discrepancy has been with us
for quite some time, and it is not obvious to find a straightforward
explanation for it. 
It is almost twice as large as the correction from the weak interactions (Section 3),
and the evaluation of the hadronic light-by-light correction would have to be off by 
a factor of 2 to 3 to explain it.

\noindent
Could it come from higher order QED effects?
After all, the coefficients $C_\mu^{(2n)}$ in Table \ref{table_1}
display a dramatic increase with $n$. An estimate of the contribution
at twelfth order $A_2^{(12)} (m_\mu / m_e )$ that shows the enhancement mechanism
mentioned previously, based on the electron ligh-by-light loop
$A_2^{(6)} (m_\mu / m_e ; {\rm LxL})$,
corrected by three electron loops, inserted in all possible ways 
in the three photon lines
that are internal to the diagram, gives \cite{QED_sum_amu}
\begin{eqnarray}
\!\!\!\!
\!\!\!\!\!\!
A_2^{(12)} (m_\mu / m_e ) \!\!\!& \sim &\!\!\! A_2^{(6)} (m_\mu / m_e ; {\rm LxL})
\left[\frac{2}{3} \ln \frac{m_\mu}{m_e} - \frac{5}{9} \right]^3 \cdot 10
\nonumber\\
\!\!\!\!
\!\!\!\!\!\!
\!\!& \sim &\!\!
0.6 \cdot 10^4
.
\end{eqnarray}
The corresponding contribution to $a_\mu$,
\begin{equation}
\delta a_\mu \sim 0.6 \cdot 10^4 \cdot \left( \frac{\alpha}{\pi} \right)^6 \sim 1 \cdot 10^{-12}, 
\end{equation}
is however way too small to explain even part of the discrepancy.

\noindent
Could higher order QCD effects be at work? Besides 
$a_{\mu}^{\rm HVP-NNLO}$ \cite{Kurz14b}
already included, higher-order corrections to $a_\mu^{\rm HLxL} $
have also been considered in \cite{Colangelo14}. The estimated value
\begin{equation}
a_\mu^{\rm HLxL ; HO} \sim 0.3(0.2) \cdot 10^{-10}
,
\end{equation}
again fails to match the size of the discrepancy.

\noindent
Finally, are we already seeing manifestations of BSM degrees of freedom? 
There are many 
proposals, see e.g. \cite{Stoeckinger10}, but the present situation
is clearly inconclusive. \\

\noindent
Clarification will hopefully come
from two new experiments, that are being prepared at FNAL \cite{FNAL,Roberts09}
and at J-PARC \cite{JPARC}, with the aim of reducing the
experimental uncertainty by a factor of 4. In order to
lead to a conclusive situation, these experimental
efforts need to be met by a comparable improvement
of the theoretical uncertainty. Forthcoming high-precision data
from e.g. VEPP-2000, BABAR, BESIII or KLOE-2, should allow for
a further reduction of the uncertainty on the hadronic
vacuum polarization part, probably below the 0.5\% level. 
Hadronic light-by-light will
then become the dominant source of theoretical errors.
There are proposals to address this issue 
\cite{Colangelo14b,Pauk14b,Colangelo14c}, but much remains to be done in
order to lead these programs to the desired goal.
Finally, increasing efforts are being made in order
to obtain reliable evaluations of the hadronic
contributions to $a_\mu$ from QCD simulations 
on the lattice. The success of this enterprise
will hinge on the ability to
develop appropriate strategies to overcome the
limitations set by the lattice. For interesting proposals
and recent advances in this direction, see \cite{Blum13}
and references therein, or the contributions in \cite{Mainz14},
as well as \cite{deRafael14} and \cite{Blum14}.
This, admittedly incomplete, list of items and of references
should merely convey the feeling that the subject remains very
active also at the theoretical level. Although the challenge is high, 
it is reasonable to expect that significant improvements
will become available around the time the data from the new experiments
will be released.\\

\noindent
I would like to close this review with a short remark
concerning the anomalous magnetic moment of the electron.
Although $a_e$ is expected to be less sensitive to new physics than $a_\mu$
by a about a factor $\sim(m_\mu/m_e)^2 \sim 40\,000$,
there are exceptions to this naive scaling, and moreover $a_e$ is
measured with a precision that is $\sim 2\,300$ times better than the 
experimental precision on $a_\mu$.
This leads one to consider the possibility \cite{Giudice12} that
degrees of freedom beyond the standard model could be first detected
in $a_e$. Indeed, improvements on both $\Delta a_e^{\mbox{\footnotesize{exp}}}$
and on the experimental determination of $\alpha$ are within
reach \cite{Terranova14} on a timescale comparable to the
new generation of $(g-2)$ experiments at FNAL and at J-PARC.

We are therefore looking forward to very exciting times
in high-precision physics with the forthcoming experiments
and, hopefully, with the accompanying improvements on
the theoretical aspects.


\section*{Acknowledgements}

\noindent
I would like to thank S. Narison and the Organizing Committee
of the QCD-2014 Conference for the invitation to give this
presentation. I have benefited from many informative and stimulating
discussions with L. Lellouch, A. Nyffeler, and E. de Rafael, and
I thank L. Lellouch and E. de Rafael for their comments on the manuscript.
I would also like to express my gratitude to the Mainz Institute for Theoretical Physics (MITP) 
for its hospitality and support. This work was supported in part by the OCEVU Labex (ANR-11-LABX-0060)
and the A*MIDEX project (ANR-11-IDEX-0001-02) funded by the “Investissements d’Avenir”
French government program managed by the ANR.


\begin{thebibliography}{99}

\bibitem{PDG12}
J. Beringer et al. [Particle Data Group], Phys. Rev. D 86, 010001 (2012).

\bibitem{Bennett06} G. W. Bennett et al.,
Phys. Rev D {\bf 73}, 072003 (2006).

\bibitem{Roberts09}
B. Lee Roberts, Chin. Phys. C {\bf 34}, 741 (2010).

\bibitem{Hanneke08}
D. Hanneke et al., Phys. Rev. Lett. {\bf 100}, 120801 (2008).

\bibitem{Schwinger48}
J. Schwinger, Phys. Rev. {\bf 73}, 416L (1948).

\bibitem{QED_calculations}
T. Kinoshita (chapter 3), 
S. Laporta and E. Remiddi (chapter 4) in {\it Lepton Dipole Moments}, 
Advanced Series on Directions in High Energy Physics -- Vol. 20,
B. Lee Roberts and William J. Marciano Eds, World Scientific Co. Pte. Ltd. (2010).

\bibitem{4and5loops}
T. Kinoshita, M. Nio, Phys. Rev. D {\bf 73}, 053007 (2006).
T. Aoyama et al., Phys. Rev. D {\bf 78}, 053005 (2008);
Phys. Rev. D {\bf 78}, 113006 (2008);
Phys. Rev. D {\bf 81}, 053009 (2010);
Phys. Rev. D {\bf 82}, 113004 (2010);
Phys. Rev. D {\bf 83}, 053002 (2011);
Phys. Rev. D {\bf 83}, 053003 (2011);
Phys. Rev. D {\bf 84}, 053003 (2011);
Phys. Rev. D {\bf 85}, 033007 (2012);

\bibitem{QED_sum_ae}
T. Aoyama et al., Phys. Rev. Lett. {\bf 109}, 111807 (2012).

\bibitem{QED_sum_amu}
T. Aoyama et al., Phys. Rev. Lett. {\bf 109}, 111808 (2012).

\bibitem{CODATA}
P. J. Mohr et al., Rev. Mod. Phys.  {\bf 84}, 1527 (2012).

\bibitem{Broadhurst93}
D. J. Broadhurst et al., Phys. Lett. B {\bf 298}, 445 (1993).

\bibitem{Laporta94}
S. Laporta, Phys. Lett. B {\bf 328}, 522 (1994).

\bibitem{Aguillar08}
P. Aguillar et al., Phys. Rev. D {\bf 77}, 093010 (2008).

\bibitem{Baikov08}
P. A. Baikov et al., Nucl. Phys. B, Proc. Suppl. {\bf 183}, 8 (2008).

\bibitem{Kurz14a}
A. Kurz et al., Nucl. Phys. B {\bf 879}, 1 (2014).

\bibitem{Bouchendira11}
R. Bouchendira et al., Phys. Rev. Lett. {\bf 106}, 080801 (2011).

\bibitem{Jackiw72}
R. Jackiw, S. Weinberg, Phys. Rev. D {\bf 5}, 2396 (1972).

\bibitem{Bars72}
I. Bars, M. Yoshimura, Phys. Rev. D {\bf 6}, 374 (1972).

\bibitem{Fujikawa72}
K. Fujikawa et al., 
Phys. Rev. D {\bf 6}, 2923 (1972).

\bibitem{Altarelli72}
G. Altarelli et al., 
Phys. Lett. B {\bf 40}, 415 (1972).

\bibitem{Bardeen72}
W. A. Bardeen et al., 
Nucl. Phys. B {\bf 46}, 315 (1972).

\bibitem{Czarnecki95}
A. Czarnecki et al., Phys. Rev. D {\bf 52}, 2619 (1995); Phys. Rev. Lett. {\bf 76}, 3267 (1996).

\bibitem{Knecht02b}
M. Knecht et al., JHEP{\bf 11}, 003 (2002).

\bibitem{Czarnecki03}
A. Czarnecki et al., Phys. Rev. D {\bf 67}, 073006 (2003); Err.-ibid. D {\bf 73}, 119901 (2006).

\bibitem{Gnendinger13}
C. Gnendinger et al., Phys. Rev. D {\bf 88}, 053005 (2013).

\bibitem{Bouchiat61}
C. Bouchiat, L. Michel, J. Phys. Radium 22, 121 (1961).

\bibitem{Durand62}
L.~Durand, Phys. Rev.  {\bf 128}, 441 (1962); Err.-ibid. {\bf 129}, 2835 (1963).

\bibitem{Gourdin69}
M. Gourdin, E. de Rafael, Nucl. Phys. B {\bf 10}, 667 (1969).

\bibitem{Davier11}
M. Davier et al., Eur. Phys. J. C {\bf 71}, 1515 (2011);
Err.-ibid. C {\bf 72}, 1874 (2012).

\bibitem{Hagiwara11}
K. Hagiwara et al., J. Phys. G {\bf 38}, 0850003 (2011).

\bibitem{Nomura13}
D. Nomura, T. Teubner, Nucl. Phys. B {\bf 867}, 236 (2013).

\bibitem{Barbieri75}
R.~Barbieri, E.~Remiddi, Nucl. Phys. B {\bf 90}, 233 (1975).

\bibitem{Krause97}
B. Krause, Phys. Lett. B {\bf 390}, 392 (1997).

\bibitem{Kurz14b}
A. Kurz et al., B {\bf 734}, 144 (2014).

\bibitem{Knecht02a}
M. Knecht, A. Nyffeler, Phys. Rev. D {\bf 65}, 073034 (2002).

\bibitem{Bijnens95}
J. Bijnens et al., Phys. Rev. Lett. {\bf 75}, 1447 (1975) [Err.-ibid. {\bf 75}, 3781 (1995)];
Nucl. Phys. B {\bf 474}, 379 (1995); Nucl. Phys. B {\bf 626}, 410 (2002).

\bibitem{Hayakawa95}
M. Hayakawa et al.,  Phys. Rev. Lett. {\bf 75}, 790 (1975);
Phys. Rev. D {\bf 54}, 3137 (1996);
Phys. Rev. D {\bf 57}, 365 (1998) [Err.-ibid. {\bf 66}, 019902(E) (2002)].

\bibitem{Prades}
J. Prades, arXiv:hep-ph/0108192.

\bibitem{tHooft74}
G 't Hooft, Nucl. Phys. B {\bf 72}, 461 (1974).

\bibitem{Witten79}
E. Witten, Nucl. Phys. B {\bf 160}, 57 (1979).

\bibitem{deRafael94}
E. de Rafael, Phys. Lett. B {\bf 322}, 239 (1994).

\bibitem{Knecht02c}
M. Knecht et al., Phys. Rev. Lett. {\bf 88}, 071802 (2002)

\bibitem{Ramsey02}
M.~J.~Ramsey-Musolf, M.~B.~Wise, Phys.\ Rev.\ Lett.\  {\bf 89}, 041601 (2002).

\bibitem{Savage92} 
  M.~J.~Savage et al.,  Phys. Lett. B {\bf 291}, 481 (1992).
  
\bibitem{Husek14}
T. Husek et al., arXiv:1405.6927 [hep-ph].

\bibitem{Pauk14}
V. Pauk, M.~Vanderhaeghen,  Eur.\ Phys.\ J.\ C {\bf 74}, 3008 (2014).

\bibitem{Mainz14}
  {\it Hadronic contributions to the muon anomalous magnetic moment Workshop. $(g-2)_{\mu}$: Quo vadis? Workshop. Mini proceedings},
  arXiv:1407.4021 [hep-ph].
  
\bibitem{Melnikov04}
K.~Melnikov, A.~Vainshtein, Phys. Rev. D {\bf 70}, 113006 (2004).

\bibitem{Nyffeler09b} 
  A.~Nyffeler, Phys.\ Rev.\ D {\bf 79}, 073012 (2009).
  
\bibitem{Prades10}
J. Prades, E. de Rafael, A. Vainshtein, chapter 9 in {\it Lepton Dipole Moments},
quoted in \cite {QED_calculations}.

\bibitem{Nyffeler09}
F. Jegerlehner, A. Nyffeler, Phys. Rept. {\bf 477}, 1 (2009);
A. Nyffeler, Phys. Rev. D {\bf 79}, 073012 (2009).

\bibitem{Colangelo14}
G. Colangelo et al., Phys. Lett. B {\bf 735}, 90 (2014).

\bibitem{Stoeckinger10}
D. St\"ockinger, chapter 12 in {\it Lepton Dipole Moments},
quoted in \cite {QED_calculations}.

\bibitem{FNAL}
R. M. Carey et al.. The New g-2 experiment - 2009. Fermilab. Proposal 0989.

\bibitem{JPARC}
  T.~Mibe [J-PARC g-2 Collaboration],
  Chin.\ Phys.\ C {\bf 34}, 745 (2010).
  
\bibitem{Colangelo14b} 
  G.~Colangelo et al., 
  JHEP {\bf 1409}, 091 (2014).
 
\bibitem{Pauk14b}
V.~Pauk, M.~Vanderhaeghen, arXiv:1409.0819 [hep-ph].

\bibitem{Colangelo14c} 
  G.~Colangelo et al., 
  Phys.\ Lett.\ B {\bf 738}, 6 (2014).
  
  \bibitem{Blum13} 
  T.~Blum et al., PoS LATTICE {\bf 2012}, 022 (2012) [arXiv:1301.2607 [hep-lat]].
  
\bibitem{deRafael14}
E. de Rafael, Phys. Lett. B {\bf 736}, 522 (2014).

\bibitem{Blum14}
T.~Blum et al., arXiv:1407.2923 [hep-lat].

\bibitem{Giudice12}
G.~F.~Giudice et al., JHEP {\bf 1211}, 113 (2012).

\bibitem{Terranova14}
F. Terranova, G. M. Tino, Phys. Rev. A {\bf 89}, 052118 (2014).


\end{thebibliography}
\end{document}